\documentstyle{mn}

\input{psfig}

\title[CLASS B0445+123: A new two-image gravitational lens system]{CLASS B0445+123: A new two-image gravitational lens system}

\author[Argo et al.]
{M.K. Argo,$^{1}$ N.J. Jackson,$^{1}$ I.W.A. Browne,$^{1}$ T. York,$^{1}$
J.P. McKean,$^{1}$ A.D. Biggs,$^{1,2}$\cr 
R.D. Blandford,$^{3}$ A.G. de Bruyn,$^{4,5}$ K.H. Chae,$^{1}$ C.D. Fassnacht,$^{6}$ L.V.E. Koopmans,$^{3}$\cr
D.R. Marlow$^{1}$, S.T. Myers,$^{7}$ M. Norbury,$^{1}$ T.J. Pearson,$^{3}$ P.M. Phillips,$^{1}$\cr A.C.S. Readhead,$^{3}$ D. Rusin$^{8,9}$
and P.N. Wilkinson$^{1}$   \\
$^{1}$University of Manchester, Jodrell Bank Observatory,
Macclesfield, Cheshire SK11~9DL, UK \\ 
$^{2}$Joint Institute for VLBI in Europe, Postbus 2, 7990 AA Dwingeloo, The Netherlands\\
$^{3}$California Institute of Technology, Pasedena, CA 91125, USA\\
$^{4}$Kapteyn Laboratory, Postbus 800, 9700 AV Groningen, The Netherlands\\
$^{5}$ASTRON, Postbus 2, 7990 AA Dwingeloo, The Netherlands\\
$^{6}$Space Telescope Science Institute, 3700 San Martin Dr., Baltimore, MD 21218, USA\\
$^{7}$National Radio Astronomy Observatory, P.O. Box 0, Socorro, NM 87801, USA\\
$^{8}$Department of Physics and Astronomy, University of Pennsylvania, 209 S. 33rd St., Philadelphia, PA 19104-6396, USA\\
$^{9}$Harvard-Smithsonian Center for Astrophysics, 60 Garden Street, Cambridge, MA 02138, USA\\
}

\date{06-05-2002}

\begin{document}
\maketitle

\begin{abstract}
A new two image gravitational lens system has been discovered as a result
of the Cosmic Lens All-Sky Survey (CLASS).  Radio observations with the 
VLA, MERLIN and the VLBA at increasingly higher resolutions all show two
components with a flux density ratio of $\sim$7:1 and a separation of 
1\farcs34.  Both components are compact and have the same spectral index. 
Follow-up observations made with the VLA at 8.4~GHz show evidence of a 
feature to the south-east of the brighter component and a corresponding 
extension of the weaker component to the north-west.  Optical observations
with the WHT show $\sim$1$\farcs7$ extended emission aligned in 
approximately the same direction as the separation between the radio 
components with an R-band magnitude of $21.8~\pm~0.4$.  
\end{abstract}


\begin{table*}
\centering
\caption{\label{obs}Radio and optical observations of B0445+123}
\begin{tabular}{cccccc}
\hline
Date		& Telescope	& Frequency/Band	& Time on source	& Resolution/Seeing	\\
(yyyymmdd)	&		&			& 			& (arcseconds)		\\
\hline
19950903	& VLA		& 8.4 GHz		& 30 seconds		& 0.2			\\
19990702	& VLA		& 15 GHz		& 3 hours		& 0.1			\\
19990903	& VLA		& 8.4 GHz		& 3 hours		& 0.2			\\
20000309	& MERLIN	& 5 GHz			& 1 hour		& 0.05			\\
20010321	& VLBA		& 5 GHz			& 1 hour		& 0.002			\\
20020203	& WHT		& R-band		& 4x600	seconds		& 1.0			\\
20011117        & Keck          & Spectrum              & 1 hour (blue), 1 hour (red)        &\\
        
\hline
\end{tabular}
\end{table*}

\begin{figure*}
\begin{tabular}{cc}
\psfig{figure=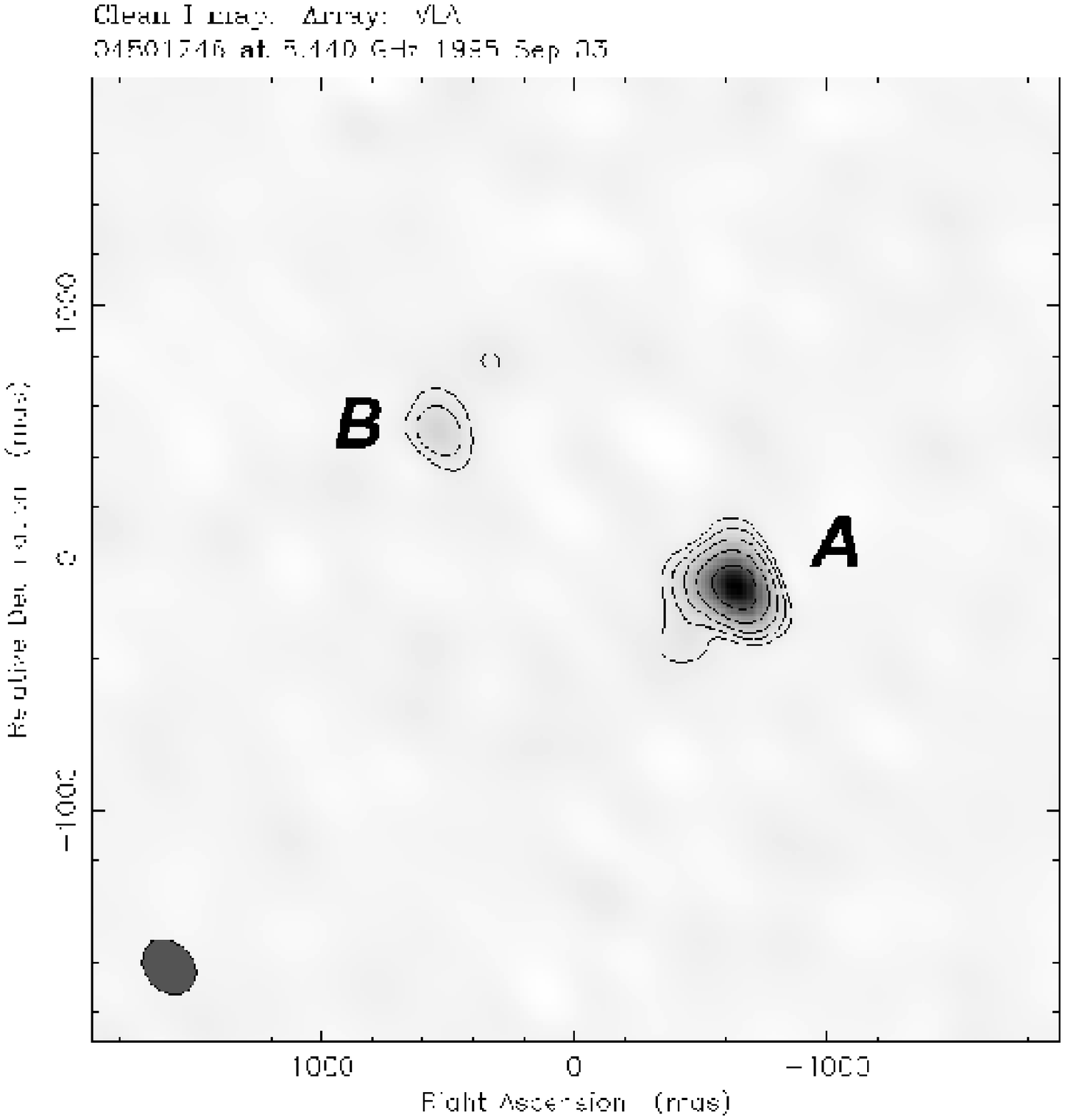,width=8.5cm}	&
\psfig{figure=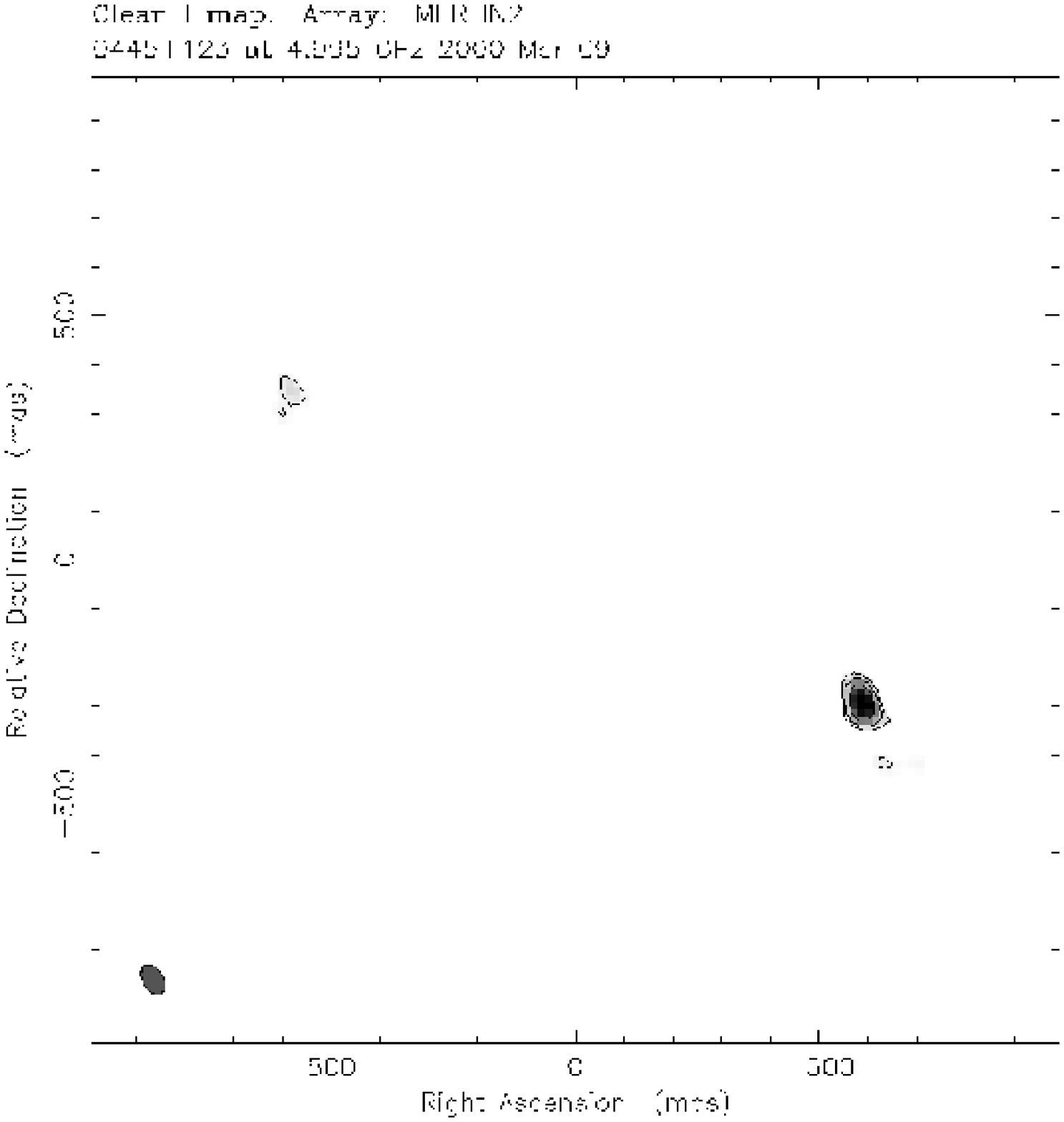,width=8.5cm}	\\
\end{tabular}
\caption{\label{vla+mer}Left: Original VLA (8.4~GHz) discovery map. Contours
are at 4, 8, 16, 32, 64 $\%$ of the peak flux density of 0.0261~Jy/beam
and the rms noise is 300~$\mu$Jy/beam. 
Right: MERLIN (5~GHz) follow-up map. Contours are plotted at 12, 24, 48,
92 $\%$ of the peak flux density of 0.0259~Jy/beam. The rms noise level
is 280~$\mu$Jy/beam.}
\end{figure*}

\section{Introduction}
The Cosmic Lens All-Sky Survey (CLASS), conducted in four phases between
1994 and 1999, has mapped a total of 16,503 radio sources with the aim of
detecting gravitationally lensed, flat-spectrum, compact objects.  The 
sources for the first two phases were selected from the 87GB (Gregory \&
Condon 1991) catalogue while those in the later CLASS-3 and 4 samples 
were chosen from the updated GB6 (Gregory et al. 1996) survey when it 
became available.  Spectral selection was performed by comparing the 
5~GHz 87GB (or GB6) flux with that from the 1.4~GHz NVSS catalogue 
(Condon et al. 1998); sources with a spectral index, $\alpha$, of 
$\leq$ -0.5 between 1.4 and 5~GHz (where $S \propto \nu^{\alpha}$) 
and total flux densities $\geq$~30 mJy at 5~GHz were then included in the
list of targets for CLASS.  The initial selection of the CLASS samples is
described in greater detail in Myers et al. (2002) and radio maps
and catalogues covering the entire CLASS sample are available at 
http://www.jb.man.ac.uk/research/gravlens/.

Each object in the survey was first mapped at a resolution of 0.2~arcseconds
with the VLA (Very Large Array) at a frequency of 8.4~GHz.  The images were
then systematically examined for signs of structure.  Potential lens 
candidates are those sources displaying multiple compact components within
15$^{\prime\prime}$ of each other.  They were selected from the CLASS sample
according to specific criteria; the full width at half maximum (FWHM) of at
least two of the components must be $\leq$~170~mas when observed with the
VLA at 200-mas resolution, the separation of the components must be 
$\geq$~300~mas, all components observed must have similar surface 
brightnesses, the sum of the 8.4-GHz flux densities must be $\geq$~20~mJy,
and the flux density ratio between the two brightest components must be 
$\leq$~10:1.

In order to distinguish between physically extended objects (e.g. core-jet
systems) and those displaying gravitational lensing effects, sources 
displaying multiple compact components in the VLA images were mapped 
at a frequency of 5~GHz and a resolution of 50~mas using MERLIN 
(Multi-Element Radio Linked Interferometer Network). Sources showing 
components with significantly different surface brightnesses at this 
stage of the survey were rejected from the search.

Sources surviving this stage of the selection process were then mapped 
at a resolution of 2~mas with the VLBA (Very Long Baseline Array). 
Browne et al. (2002) describes the candidate selection criteria
and follow up procedures in greater detail.

In this paper we describe the follow up observations of one of these 
candidates, CLASS B0445+123, which lead us to the conclusion that it 
is a two-image gravitational lens system.

\section{Radio Observations}
CLASS B0445+123 was first imaged with the VLA in 1995 during the second 
phase of CLASS and follow-up observations were carried out with both the
VLA, MERLIN and the VLBA.  The data were calibrated using AIPS (Astronomical
Image Processing System, distributed by the U.S. National Radio Astronomy
Observatory) and then mapped and self-calibrated in DIFMAP (Shepherd 1997). 
A summary of all the observations made of B0445+123 is given in Table 
\ref{obs} and the flux densities of both components for each observation
are listed in Table \ref{components}.

\begin{table*}
\centering
\caption{\label{components}The two components of B0445+123.  All flux 
densities are given in mJy and are accurate to $\simeq$5\%.  
V95 indicates the original VLA observation 
made in 1995, while V99 indicates the follow-up observations made with 
the VLA in 1999.  The dates of each observation are given in Table \ref{obs}.}
\begin{tabular}{cccccccc}
\hline
Component	& RA 		& Dec 		& S$_{5}$	& S$_{5}$	& S$_{8.4}$	& S$_{8.4}$	& S$_{15}$	\\
		& (J2000)	& (J2000)	& (MERLIN)	& (VLBA)	& (V95)		&  (V99)	& (V99)		\\
\hline
A		& 04 48 21.990	& +12 27 55.409	& 25.0		& 16.6		& 27.1		&  24.5		& 32.0		\\
B		& 04 48 22.070	& +12 27 56.018	& 4.2		& 2.7		& 3.7		&  4.4		& 4.3		\\
\hline
\end{tabular}
\end{table*}

\begin{figure*}
\begin{tabular}{cc}
\psfig{figure=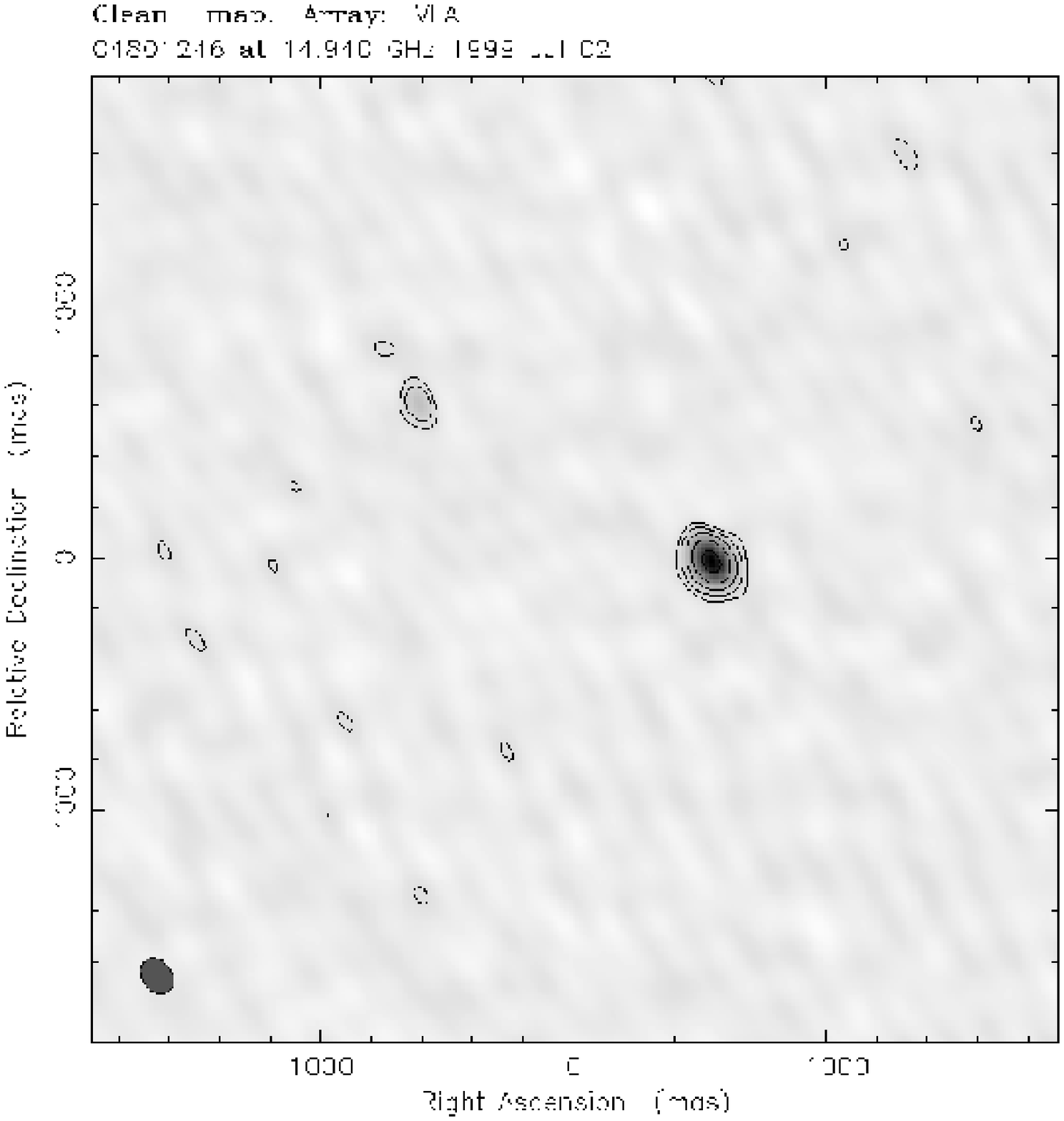,width=8.5cm}&     
\psfig{figure=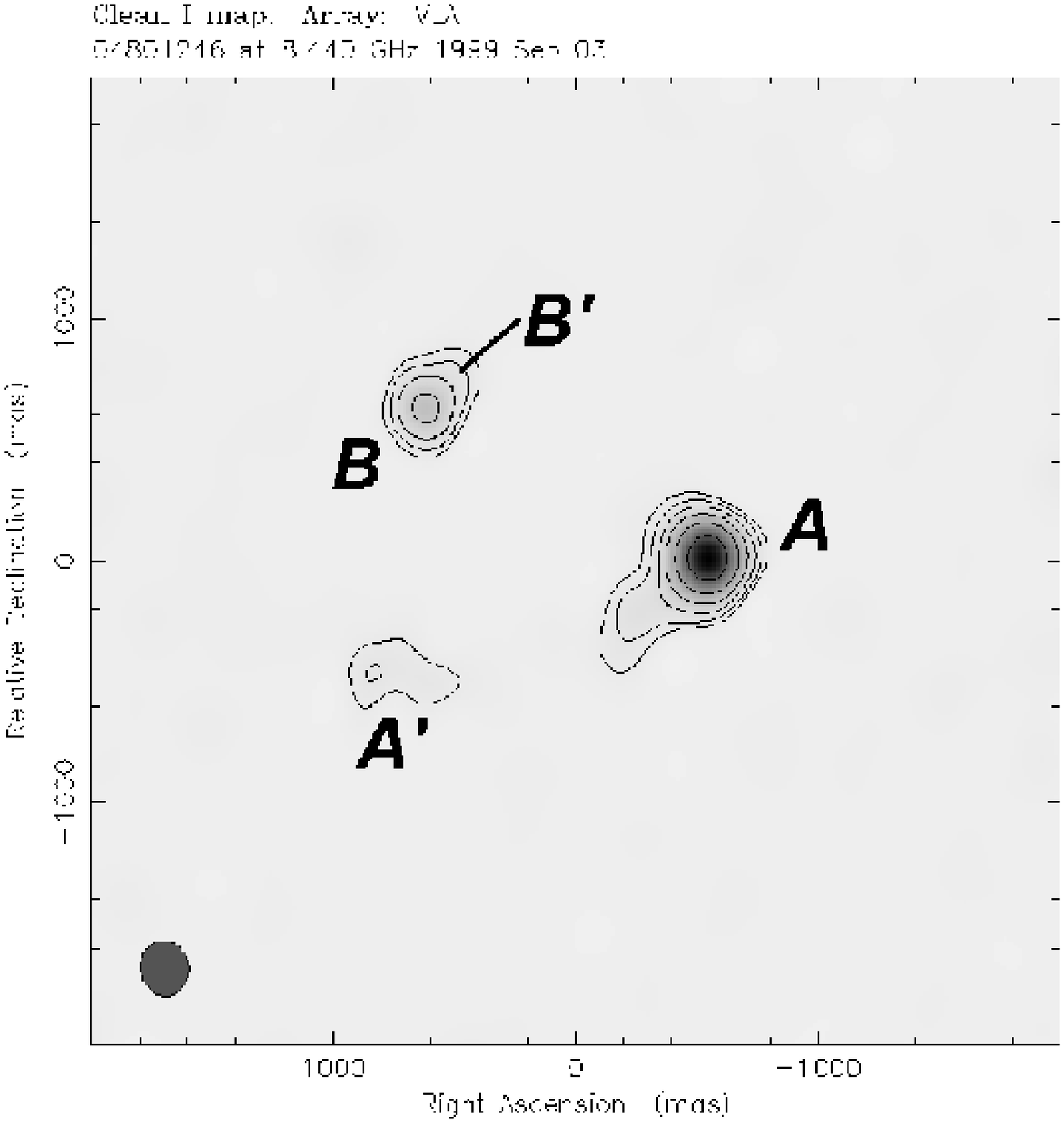,width=8.5cm}\\
\psfig{figure=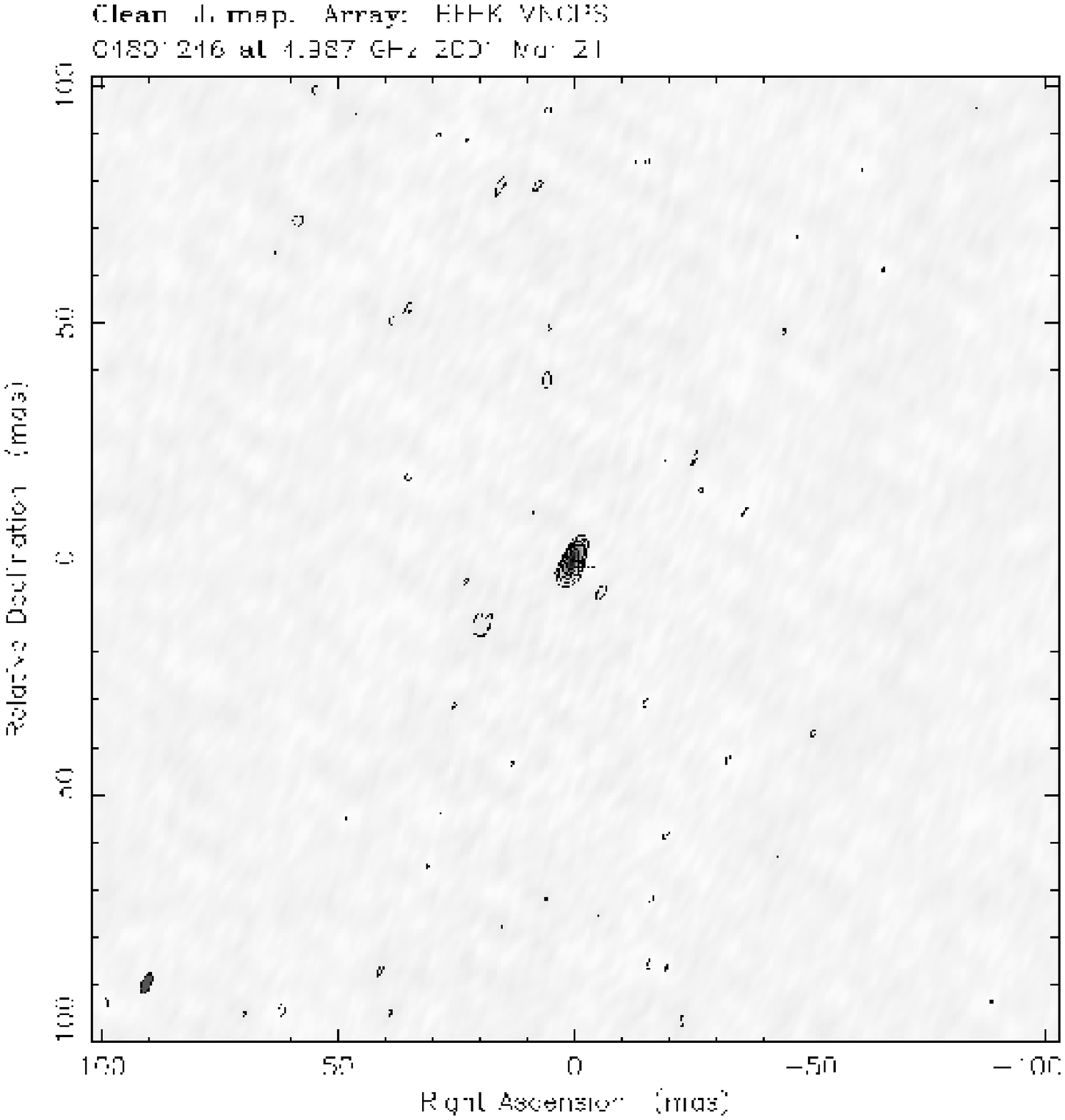,width=8.5cm}&
\psfig{figure=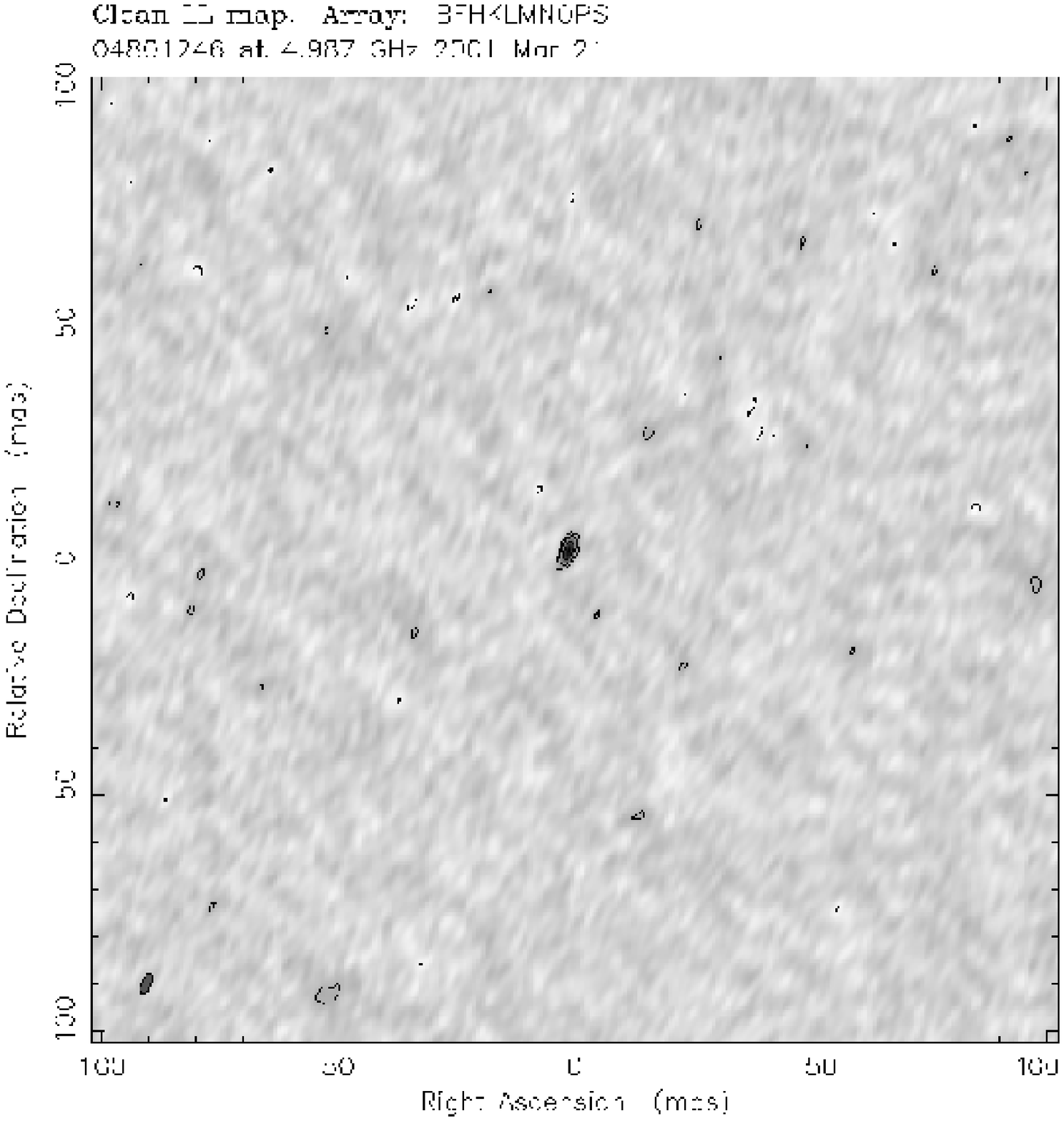,width=8.5cm}\\    
\end{tabular}
\caption{\label{vla+vlba}Top left: VLA 15 GHz follow-up map; contours are
plotted at 5, 10, 20, 40, 80 $\%$ of the peak flux density of 0.0284 Jy/beam.
Top right: VLA 8.4 GHz follow-up map; contours are at 2, 4, 8, 16, 32, 
64 $\%$ of the peak flux density of 0.0225 Jy/beam; 
$\sigma_{rms}\simeq20\mu$Jy/beam. Bottom: VLBA (5 GHz)
snapshots of component A (left) and B (right), contours are plotted at -3,
3, 6, 12, 24, 48 $\times \sigma_{\rm rms}\simeq 170 \mu$Jy/beam for A and -3, 
3, 6, 12 $\times \sigma_{rms}\simeq 140 \mu$Jy/beam for B.}
\end{figure*}

\subsection{VLA}
B0445+123 was observed with the VLA in its largest, ``A" configuration on 03
September 1995.  Myers et al. (2002) describe the VLA observations and the processing
carried out on the CLASS data in greater detail.  The original VLA image
(Figure \ref{vla+mer}, left) shows two distinct components with a separation
of 1\farcs34 and a flux density ratio of 7.3:1.  The brighter and fainter
components are given the designations A and B respectively, according to
the usual convention, and they have a combined flux density at 8.4~GHz of
30.8~mJy.   Component A appears slightly extended towards the south-east.

Further observations were made in 1999 at 15 and 8.4~GHz (Figure 
\ref{vla+vlba}, top left and top right respectively).  Deep observations
at 8.4~GHz show a clear extension to the south-east of component A,
including a weaker component A$^{\prime}$, and 
signs of a corresponding elongation (B$^{\prime}$) 
of component B to the north-west.

\subsection{MERLIN}
Snapshot observations of B0445+123 were made on 09 March 2000 with MERLIN
at a frequency of 5~GHz and a resolution of 50~mas.  The data were phase
calibrated using the source B0441+106 with a switching cycle of three 
minutes on source to two minutes on the calibrator.  Figure \ref{vla+mer},
top right, shows the MERLIN map after self-calibration in DIFMAP. 
The two components seen in the VLA images were detected.

\subsection{VLBA}
The two components of B0445+123 were imaged individually at 5~GHz with the
VLBA at a resolution of 2~mas on 21 March 2001.  Figure \ref{vla+vlba}, 
bottom left and bottom right, show the VLBA maps of components A and B 
respectively.  The observations show no structure or extension to either
component and give a separation of 1340 mas. The 5~GHz flux densities of
the components measured with the VLBA are considerably lower than that 
measured with MERLIN.  This could be because either the source has low 
surface brightness emission which is not detected in the high resolution
VLBA observations, or because the source is variable.  More observations
of B0445+123 are necessary in order to determine which is the case.

\section{Optical Observations}
The APM (Automated Plate Measuring machine, Irwin et al. 1994) catalogue 
obtained from Palomar Observatory Sky Survey 1 shows no associated 
optical components
at the position of B0445+123 indicating that any counterpart must have a
blue magnitude of $\geq$~21 (the limiting magnitude detected by the APM).

\begin{figure}
\psfig{figure=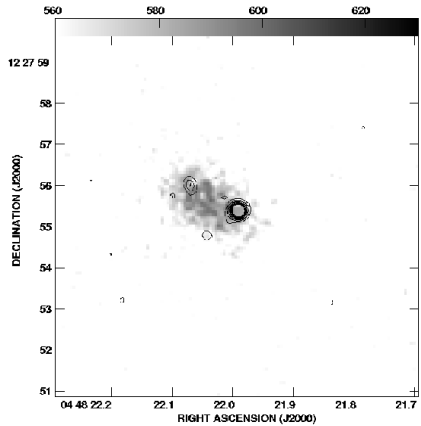,width=8cm,angle=-90}
\caption{\label{wht}Contours from the VLA discovery map overlayed on the optical picture obtained with the WHT.  The contours are plotted at -10 to +10 (in steps of 1) $\times$ 1 mJy/beam.  The line AB is at an angle of $62.5^\circ$ from north.  The radio/optical registration is believed to be accurate to $\pm$0\farcs3}
\end{figure}

An optical image in R-band was obtained with the WHT (William Herschel 
Telescope) on La Palma during service observations.  The observing 
conditions were photometric and the seeing was 1.0~arcsec. The observations
show a fuzzy blob approximately 1\farcs7 in diameter at the position of the
radio components and extended in the same orientation.  Figure \ref{wht}
shows the WHT image of B0445+123 with contours from the original 8.4~GHz
VLA observations overlayed. Astrometric calibration was performed in 
GAIA\footnote{Part of the Starlink Project which is run by CCLRC on 
behalf of PPARC.} using the seven objects in the field also detected 
on the APM.  The random astrometric error is $\pm$0\farcs3 comparable 
to the $\sim$0\farcs5 offset from the radio reference frame.  By 
comparison with Landolt standard star 93-407 (Landolt 1992) the object 
was found to have an R-band magnitude of $21.8~\pm~0.4$.  

An optical spectrum was obtained with the 10-m Keck Telescope and the
LRIS spectrograph on 17 November
2001.  Strong continuum emission was present, with a redshift of 0.557
resulting from detections of Ca H,K and Mgb lines but there are no obvious 
emission lines which might indicate a redshift for a lensed object 
(McKean et al. 2002, in preparation).

\section{Discussion: is CLASS B0445+123 a lens system?}

\begin{figure}
\psfig{figure=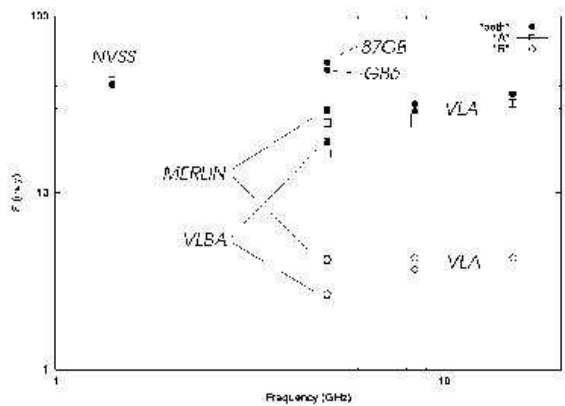,width=8.5cm}
\caption{\label{fluxes}Radio spectra of the two components of B0445+123. 
Note that the 5 GHz MERLIN and VLBA observations have very different 
resolutions and at least some of the difference in flux densities results
from extended emission being resolved out in the VLBA observations.}
\end{figure}

The radio spectra for the components (Figure \ref{fluxes}) confirm
that both have similar spectral indices as required by the lensing hypothesis.
The presence of two or more components separated by $\sim$1~arcsec, each
compact on milliarcsecond scales and with near identical radio spectra 
(Figure \ref{fluxes}) has so far proven to be a 100\% reliable 
indicator of gravitational lensing within CLASS (Browne et al. 2002).
Only one system with two components with milliarcsecond-scale structure 
is thought to be a binary quasar - CLASS B0827+525 (Koopmans et al. 2000).

The fact that the optical object seen in the WHT image is extended 
in the same orientation as the radio components and is larger than 
the separation of the two radio components suggests that some emission
from both a lensing galaxy and at least one image is being detected.
This is also consistent with the lensing hypothesis. If the object were a
binary quasar, optical emission would be expected to be centred on one or
other (or both) of the radio point sources.

From the above arguments, we conclude that CLASS B0445+123 is highly
likely to be a gravitationally lensed system. On this assumption, 
we now try to produce a
plausible lens model for the structure seen in the 8.4~GHz radio map
(Fig. \ref{vla+vlba}).

\begin{figure*}
\begin{tabular}{cc}
\psfig{figure=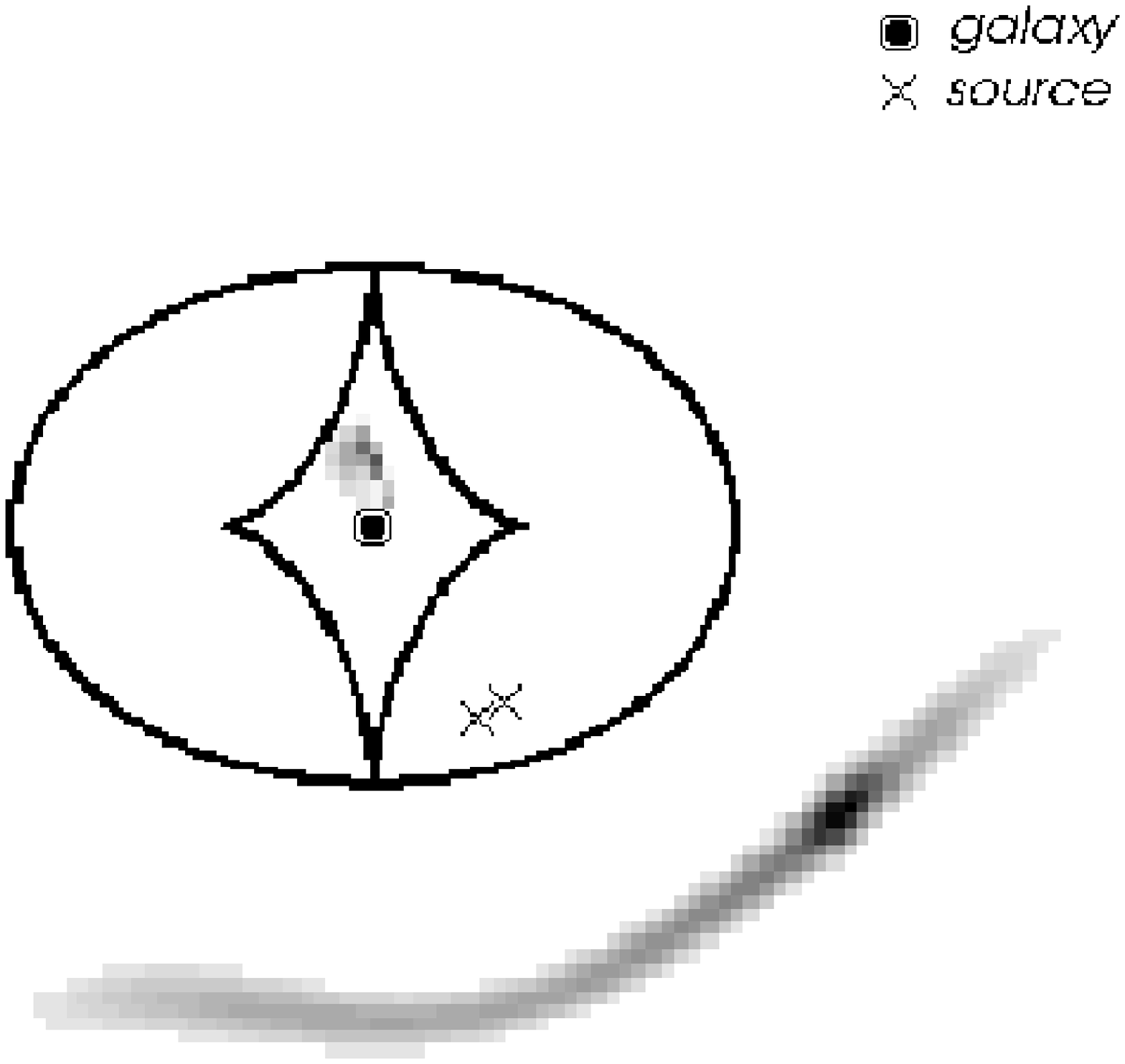,width=8.5cm}&     
\psfig{figure=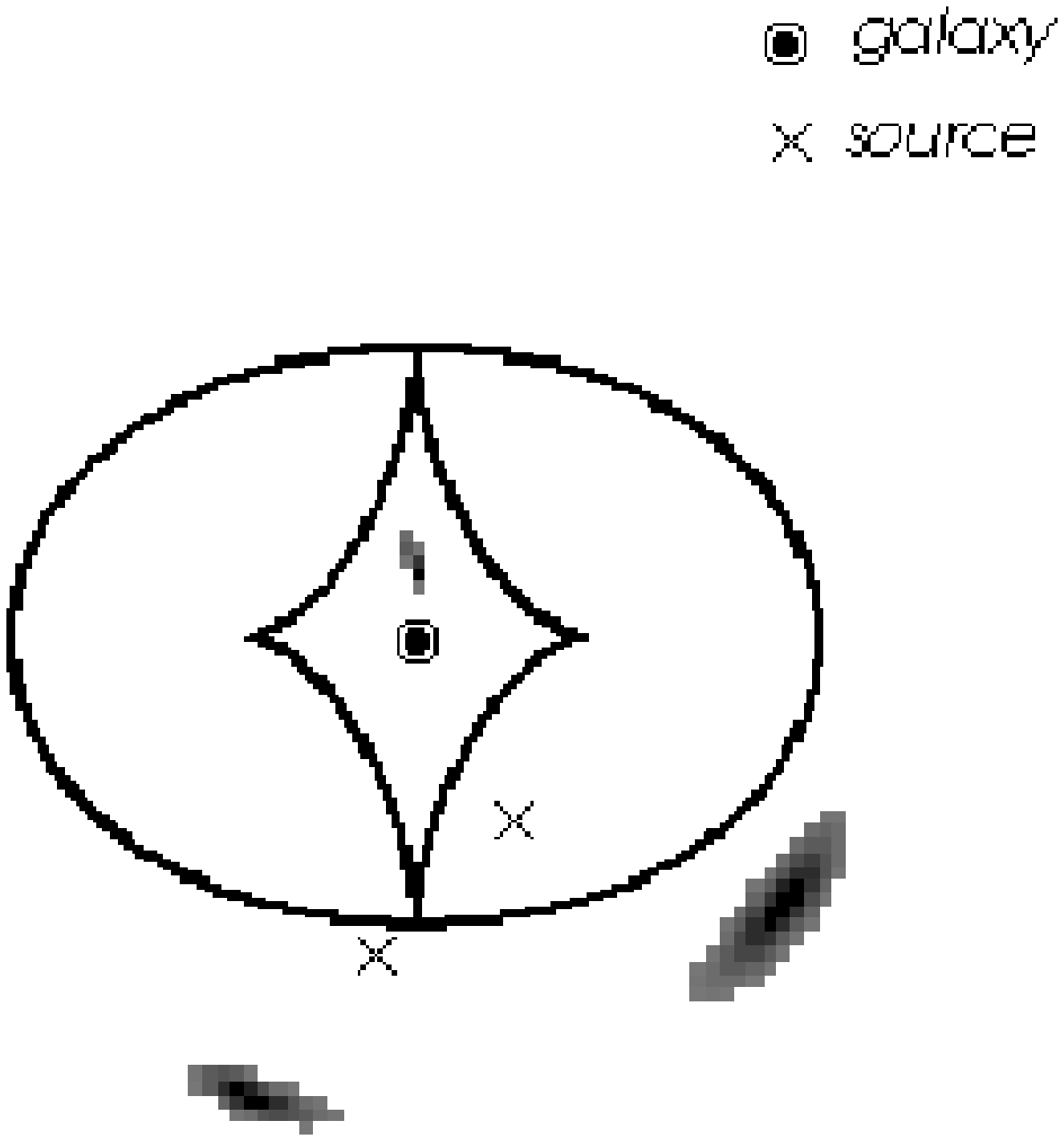,width=8.5cm}\\
\end{tabular}
\caption{\label{models}Simple models for B0445+123.  In the model on the
left, both core and jet lie within the outer caustic, while in the model
on the right, the jet lies outside the caustic such that only the core is
doubly imaged.}
\end{figure*}

Figure \ref{models} shows two possible simple mass 
models for B0445+123.  Each one shows the positions of the lensing galaxy
and the background sources with the caustics and image positions overlaid. 
In both models the background source has two components, a compact core 
and jet. They differ in that in the first both core and jet lie within the
outer caustic and thus both are doubly imaged, while in the second, only
the core is doubly imaged.  To first order, both models reproduce the 
structure seen in the 8.4-GHz follow-up maps (see Figure \ref{vla+vlba}).
The fact that the first model can reproduce the slight extension to the 
north-west of component B suggests that this is the more likely of the two.

Although preliminary lens models of the system can reproduce the radio 
structure, we emphasize that, because of the dearth of observational
constraints, we are only using the modelling to show that the lens hypothesis
is plausible.  However, with deep MERLIN observations at L-band we would
expect to see much more of the extended substructure in the images. This,
together with a high resolution optical image to tie down the relative 
astrometry of the lens and the images, would mean that B0445+123 has the
potential to be an interesting system for determination of an accurate 
lens model and in which to search for the effects of the sub-structure 
in the lens predicted by CDM galaxy formation scenarios (e.g. 
Metcalf, 2002).

\smallskip
\noindent {\large \bf ACKNOWLEDGMENTS}
\smallskip

MERLIN is run by the University of Manchester as a National Facility on 
behalf of PPARC. 
The WHT is operated on the island of La Palma by the Isaac 
Newton Group in the Spanish Observatorio del Roque de los 
Muchachos of the Instituto de Astrof\'{\i}sica de Canarias 
The observations made with the WHT were carried out as part 
of the ING Service Programme. The VLA and the VLBA are operated by the 
National Radio Astronomy Observatory which is a facility of the National
Science Foundation operated under cooperative agreement by Associated 
Universities, Inc. We thank the W.M. Keck foundation for the generous 
grant that made the W.M. Keck Observatory possible. MA acknowledges
support from the summer student programme at Jodrell Bank Observatory,
University of Manchester. JPM, MN, PMP and TY
thank PPARC for support in the form of postgraduate studentships.

\smallskip
\noindent {\bf REFERENCES}
\smallskip

\noindent Browne I.W.A. et al., 2002, MNRAS, submitted\newline
Condon J.J. et al., 1998, AJ, 115, 1693\newline
Gregory P.C., Condon J.J., 1991, ApJS, 75, 1011\newline
Gregory P.C. et al., 1996, ApJS, 97, 347\newline
Irwin M.J., Maddox S.J., McMahon R.G., 1994, Spectrum (Newsletter of the
Royal Greenwich Observatory), 2, 14\newline
Koopmans L.V.E. et al., 2000, A\&A, 361, 815\newline
Landolt A.U., 1992, AJ, 104, 340\newline
McKean J. et al., 2002, in preparation\newline
Metcalf R.B., 2002, astro-ph/0203012\newline
Myers S.T. et al., 2002, MNRAS, submitted\newline
Shepherd M.C., 1997, Astronomical Data Analysis Software and Systems VI, ASP Conference Series, Vol. 125, p 77.\newline

%location of psfig.tex in afile2

\end{document}